\documentclass[sigconf,natbib=true,nonacm=true]{acmart}

\usepackage{amsmath}

\usepackage{graphicx}
\usepackage{xcolor}
\usepackage{booktabs}
\usepackage{array}
\usepackage{multirow}
\usepackage{threeparttable}
\usepackage{tabularx}
\usepackage{makecell}

\usepackage{xurl} 

\usepackage{adjustbox}
\usepackage{subcaption}
\usepackage{float}
\usepackage{placeins}

\usepackage{enumitem}
\usepackage[breakable,skins]{tcolorbox}
\tcbuselibrary{skins,breakable}

\definecolor{gaincolor}{RGB}{25,120,65}
\definecolor{losscolor}{RGB}{190,45,45}

\providecommand{\eg}{\emph{e.g., }}

\newcommand{\myitem}[1]{\noindent$\bullet$ \textbf{#1.} }

\newtcolorbox{casestudy}[1]{
    enhanced,
    breakable,
    title={#1},
    colback=black!1,
    colframe=blue!45!black,
    colbacktitle=blue!7!white,
    coltitle=blue!45!black,
    fonttitle=\bfseries,
    boxrule=0.4pt,
    borderline west={1.8pt}{0pt}{blue!60!black},
    arc=1mm,
    left=6pt,
    right=6pt,
    top=5pt,
    bottom=5pt,
    before skip=8pt,
    after skip=8pt
}

\AtBeginDocument{%
}

\setcopyright{none}
\begin{document}

\title{AutoRSE: Evaluating LLM Agents as Recommendation Engineers
for Self-Evolving Recommender Systems}

\settopmatter{authorsperrow=1}

\author{\texorpdfstring{\large
Rui Tian\textsuperscript{1,*},
Yi Fang\textsuperscript{1,*},
Yejin Hua\textsuperscript{1},
Du Wang\textsuperscript{1},
Yuzhe Wang\textsuperscript{1},
Minxuan Shan\textsuperscript{1}\\[2pt]
Zirui Ren\textsuperscript{1},
Xinyu Lin\textsuperscript{2},
Chongming Gao\textsuperscript{1},
Wenjie Wang\textsuperscript{1,\ensuremath{\dagger}},
Fuli Feng\textsuperscript{1}
}{Rui Tian, Yi Fang, Yejin Hua, Du Wang, Yuzhe Wang, Minxuan Shan, Zirui Ren, Xinyu Lin, Chongming Gao, Wenjie Wang, Fuli Feng}}

\affiliation[obeypunctuation=true]{
    \institution{\textsuperscript{1}University of Science and Technology of China},
    \city{Hefei}, \country{China}\\
    \institution{\textsuperscript{2}National University of Singapore},
    \country{Singapore}\\[3pt]
}

\renewcommand{\shortauthors}{Tian et al.}

\begin{abstract}
Recommender Systems Engineer Agents (RSE-Agents), which use LLM-based agents to autonomously improve recommendation algorithms, are emerging as a promising paradigm. However, their capabilities and limitations remain poorly understood due to the lack of a unified, standardized evaluation framework and systematic trajectory-level analysis. To address this gap, we introduce AutoRSE, an evaluation framework spanning six representative industrial-grade recommendation tasks, and conduct a large-scale empirical study of three frontier LLMs across three widely used agent harnesses.
Our results show that RSE-Agents consistently enhance the performance of backbones on five of the six tasks, with an average relative improvement of 16.1\%, indicating that they can already autonomously improve recommendation performance. However, trajectory-level analysis also reveals that current RSE-Agents function primarily as engineering optimizers rather than algorithmic improvers: their gains are largely driven by hyperparameter tuning, training strategies, and model ensembling, while genuine algorithmic innovations are rare. Since the largest gains in recommender systems often stem from algorithmic advances, this behavior underscores a fundamental limitation of current RSE-Agents.
To probe this limitation, we perform idea-injection experiments and show that agents equipped with innovative ideas from frontier research or human experts achieve substantially larger improvements than those relying on self-generated ideas. These results suggest that idea generation is a central bottleneck of current RSE-Agents, highlights a key direction for the future recommendation system research. Our implementation is publicly available on GitHub: \textcolor{blue!50!black}{\url{https://github.com/Ray-Tiann/AutoRSE}}.
\end{abstract}


\maketitle

\begingroup
\renewcommand{\thefootnote}{\fnsymbol{footnote}}
\footnotetext[1]{Equal contribution.}
\footnotetext[2]{Corresponding author.}
\endgroup

\raggedbottom

\section{Introduction}

\begin{figure}[t]
    \centering
    \includegraphics[width=\linewidth]{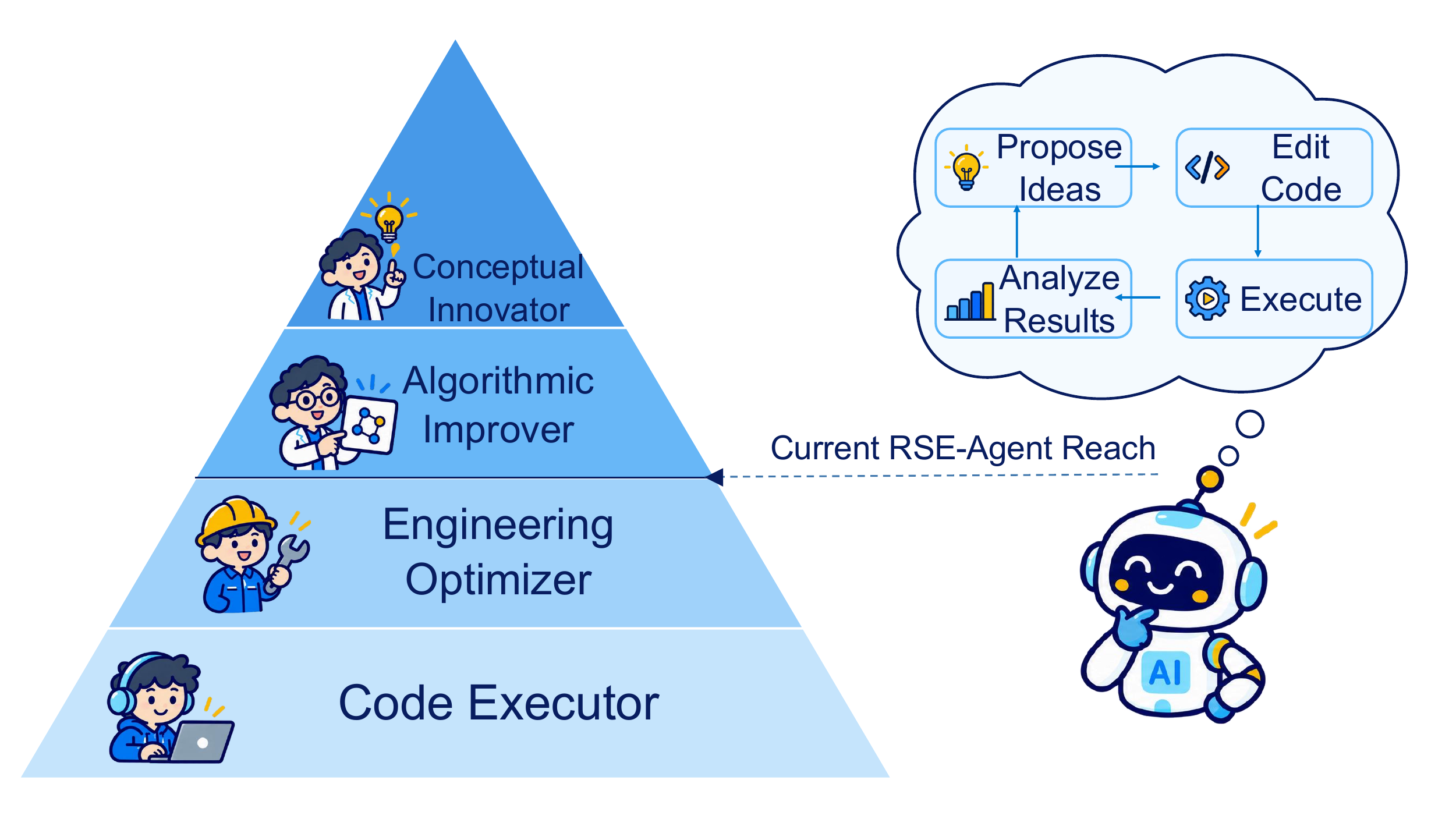}
    \caption{A four-level hierarchy of research capabilities for RSE-Agents: current LLM agents achieve most recommendation improvements through engineering-level refinements, while higher-level idea generation remains rare.}
    \label{fig:rse-agent-pyramid}
\end{figure}


Large language model (LLM) agents, such as Claude Code~\citep{anthropic_claudecode} and Codex~\citep{openai_codex}, have recently demonstrated strong capabilities on long-horizon, real-world engineering tasks, such as software engineering~\citep{yang2024sweagent} and machine learning engineering~\citep{chan2024mlebench}. 
Inspired by these advances, Recommender Systems Engineer Agents (RSE-Agents) have emerged as a promising direction for automating recommender system development~\citep{lao2026agentx,wang2026selfevolving}. 
Specifically, RSE-Agents aim to use LLM-based agents to autonomously improve recommendation algorithms by proposing new ideas, implementing them in code, and validating them through experiments.
By taking over the role of human recommender systems engineers, RSE-Agents have the potential to significantly accelerate recommendation algorithm development while reducing engineering cost.

However, research on RSE-Agents is still in its early stages. Existing pioneering systems, such as AgentX~\citep{lao2026agentx} and Self-Evolving Recommendation System~\citep{wang2026selfevolving}, mainly demonstrate the practical feasibility of agent-driven recommendation iteration, each within its own proprietary industrial pipeline. As a result, they provide neither a general framework for quantitatively evaluating and comparing different RSE-Agents, nor a systematic understanding of how such agents actually achieve performance gains. Consequently, two fundamental questions remain open: (i) how should we evaluate RSE-Agents in a unified and realistic manner, and (ii) what are the capability boundaries and key bottlenecks of current LLM-based agents on this task?

To answer these questions, we conduct a large-scale empirical study of RSE-Agents, with an overview presented in Figure~\ref{fig:agent_workflow}. We first build AutoRSE, an evaluation framework spanning six representative industrial-grade recommendation tasks, covering both generative recommendation~\citep{zhai2024hstu,yan2025intsr,rajput2023tiger} and click-through rate (CTR) prediction~\citep{zhang2025onetrans,zhu2025rankmixer,zhang2024wukong}. For each task, we provide the LLM agent with a faithfully reproduced state-of-the-art algorithm, a sandboxed environment with one H200 GPU, and a 12-hour optimization budget, and ask it to autonomously improve the baseline. Within this framework, we evaluate three frontier LLMs (Claude Opus 4.8~\citep{anthropic2026opus48}, GPT-5.5~\citep{openai2026gpt55}, and DeepSeek V4 Flash~\citep{deepseek2026v4}) with three widely used agent harnesses (Claude Code, Codex, and Terminus-2~\citep{harbor_terminus2}). Across five of the six tasks, RSE-Agents consistently improve upon strong baselines, achieving an average relative improvement of 16.1\%. These results suggest that current RSE-Agents can already autonomously improve recommendation performance.

However, a deeper trajectory-level analysis reveals a less encouraging picture. Although RSE-Agents are effective at improving performance, they rarely improve the core recommendation algorithms themselves. Instead, current agents mostly act as \textbf{Engineering Optimizers} rather than \textbf{Algorithmic Improvers} (Figure~\ref{fig:rse-agent-pyramid}): most of their gains come from hyperparameter tuning, training strategies, and model ensembling, while genuine algorithmic innovations are rare. This distinction matters because the historically largest performance gains in recommender systems typically come from algorithmic advances rather than engineering refinements alone~\citep{zhai2024hstu,zhang2024wukong,rajput2023tiger}. Thus, this reliance on engineering tuning underscores a fundamental limitation of current RSE-Agents.

To further investigate this limitation, we conduct idea-injection experiments. We find that when agents are supplied with innovative ideas drawn from frontier research or human experts, they achieve substantially larger improvements than when relying solely on self-generated ideas, surpassing the latter by 13.35\%. This result suggests that idea generation, rather than implementation or experimental execution, is the key bottleneck of current RSE-Agents. More broadly, it points to a promising direction for future research: equipping agents with stronger capabilities for generating novel algorithmic ideas.
In summary, this work makes three main contributions:

\setlength{\leftmargini}{10pt} \begin{itemize}
    \item We build a unified evaluation framework for RSE-Agents, covering six representative industrial-grade recommendation tasks, and benchmark frontier LLMs with widely used agent harnesses.
    \item Through Trajectory-level analysis, we find that current agents cleanly master the lower tiers of execution and engineering optimization, yet exhibit only marginal capabilities in algorithmic innovation and remain far from conceptual innovation, as shown in Figure~\ref{fig:rse-agent-pyramid}.
    \item Through idea-injection experiments, we show that innovative ideas from frontier research and human experts lead to much larger improvements than agents' self-generated ideas, identifying idea generation as the key bottleneck of current RSE-Agents and a promising direction for future progress.
    \item We release our benchmark code and agent trajectories at \textcolor{blue!50!black}{\url{https://github.com/Ray-Tiann/AutoRSE}}, alongside a live leaderboard supporting open submissions to continuously evaluate emerging RSE-Agents.
\end{itemize}

\section{Method}

In this section, we introduce AutoRSE(as shown in figure~\ref{fig:agent_workflow}), a unified evaluation framework for assessing whether RSE-Agents can autonomously improve recommendation algorithms. Specifically, we first formulate the RSE task (\S\ref{sec:task_def}), then detail the AutoRSE evaluation framework—covering the recommendation task suite, execution environment, and evaluation protocol (\S\ref{sec:framework})—and finally present our trajectory-level analysis methodology (\S\ref{sec:trajectory-analysis}).

\subsection{RSE Task Formulation}
\label{sec:task_def}
AutoRSE evaluates the ability of LLM agents to autonomously improve existing recommendation baselines across diverse scenarios. Specifically, each RSE task instance is formulated as a tuple $(\mathcal{I}, \pi_0, \mathcal{D}, B)$, consisting of a task description $\mathcal{I}$, an initial recommendation baseline $\pi_0$, a dataset $\mathcal{D}$, and a computational budget $B$. The task description $\mathcal{I}$ specifies the recommendation objective, the primary evaluation metric $\mathcal{M}$, and the submission requirements. The initial baseline $\pi_0$ is provided as a complete reference codebase implementing the full model architecture and training pipeline. The dataset $\mathcal{D}$ provides training data, validation data, and unlabeled test inputs to the agent, whereas ground-truth test labels $Y_{\mathrm{test}}$ remain strictly hidden for final evaluation.

\begin{figure*}[t]
    \centering
    \includegraphics[width=\textwidth]{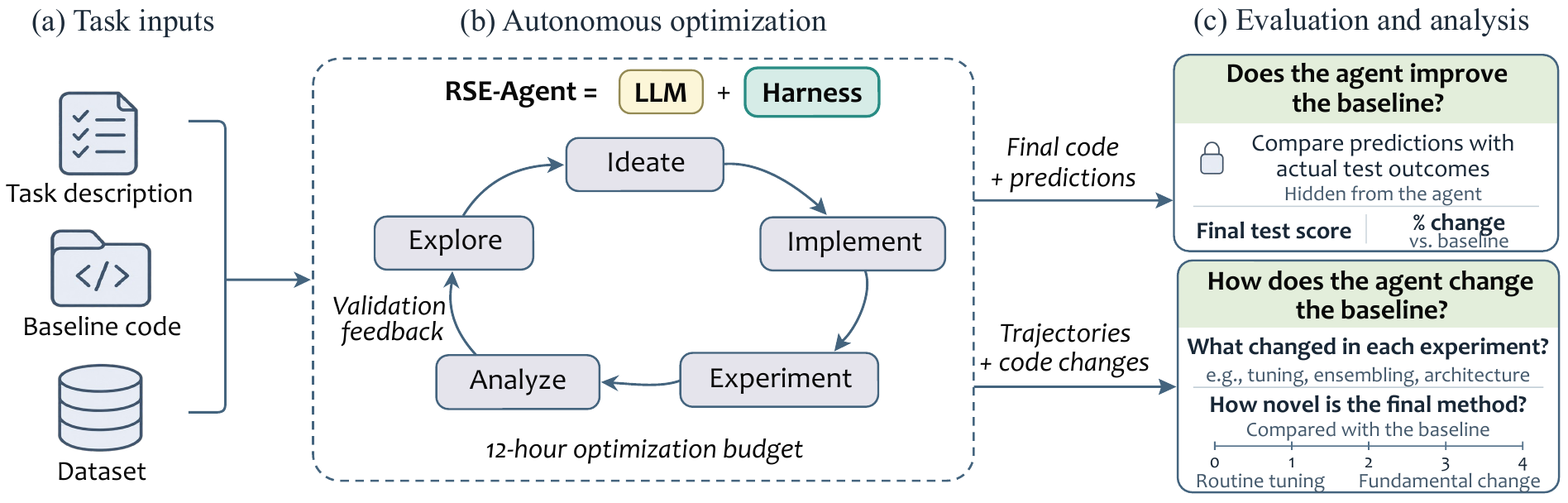}
    \caption{Overview of AutoRSE: (a) task inputs; (b) autonomous baseline optimization by an LLM and harness within a 12-hour budget; and (c) held-out prediction evaluation, strategy classification, and novelty assessment using trajectories and code changes.}
    \label{fig:agent_workflow}
\end{figure*}
As illustrated in Figure~\ref{fig:agent_workflow}, solving an RSE task follows an iterative five-stage loop: \textbf{Explore}, \textbf{Ideate}, \textbf{Implement}, \textbf{Experiment}, and \textbf{Analyze}. Let $\pi_i$ denote the recommendation codebase at iteration $i$ (initialized as the baseline $\pi_0$). In each cycle, the agent explores $\pi_i$, proposes a hypothesis, and implements code modifications, which are then executed on dataset $\mathcal{D}$ to obtain environment feedback $\text{Feedback}(\pi_i, \mathcal{D})$ (i.e., stdout logs, error traces, and validation metrics). Formally, the agent analyzes this feedback to update the codebase state:
$$
\pi_{i+1} = \text{AgentUpdate}\left(\pi_i, \text{Feedback}(\pi_i, \mathcal{D})\right),
$$
determining whether to retain, revise, or discard the modification. Repeating this loop until the computational budget $B$ is exhausted yields the best-validated strategy $\pi_{\text{final}}$ and its test predictions $\hat{Y}_{\text{final}}$.
\subsection{The AutoRSE Evaluation Framework}
\label{sec:framework}

\myitem{Recommendation Task Suite}
AutoRSE provides a general evaluation interface: in principle, any recommendation problem can be incorporated into the framework once instantiated as the tuple $(\mathcal{I}, \pi_0, \mathcal{D}, B)$. In this paper, we instantiate AutoRSE with six representative tasks, as summarized in Table~\ref{tab:tasks}. We focus on six tasks because industrial-grade recommendation algorithms are challenging to reproduce faithfully\footnote{We rigorously verify the correctness of our reproduced baselines against the original papers; see Appendix Section~\ref{app:task_descriptions} for reproduction details.} and large-scale evaluation of LLM agents on many such tasks incurs substantial computational cost.

Specifically, our task selection follows two criteria. First, the tasks should be \textbf{industrial-grade}: they target core recommendation problems in modern large-scale systems, especially generative recommendation and click-through rate prediction. Second, the tasks should be \textbf{representative}: within these problem settings, we select influential algorithms with distinct modeling strategies, including sequential modeling (HSTU)~\cite{zhai2024hstu},
semantic-ID generation (TIGER)~\cite{rajput2023tiger},
joint search--recommendation modeling
(IntSR)~\cite{yan2025intsr},
feature interaction (WuKong)~\cite{zhang2024wukong},
multi-head token mixing (RankMixer)~\cite{zhu2025rankmixer},
and unified sequence modeling
(OneTrans)~\cite{zhang2025onetrans}.
Together, these six tasks provide a diverse and practically grounded test suite for evaluating whether LLM agents can improve recommendation baselines across major industrial recommendation scenarios.

\begin{table}[t]
\centering
\small
\caption{Recommendation tasks in AutoRSE.}
\label{tab:tasks}
\setlength{\tabcolsep}{2pt}
\renewcommand{\arraystretch}{1.15}

\begin{tabularx}{\columnwidth}{
    @{}ll>{\raggedright\arraybackslash}Xl@{}
}
\toprule
Type & Task & Dataset & Metric ($\mathcal{M}$) \\
\midrule
\multirow{3}{*}{
    \shortstack[l]{Generative\\recommendation}
}
& HSTU  & Amazon Beauty & NDCG@10 \\
& TIGER & Amazon Beauty & NDCG@10 \\
& IntSR & Amazon Kindle
& \shortstack[l]{Macro NDCG@10} \\
\midrule
\multirow{3}{*}{
    \shortstack[l]{Click-through\\rate prediction}
}
& WuKong    & MovieLens Latest & AUC \\
& RankMixer & Criteo 5M        & AUC \\
& OneTrans  & KuaiSAR v2       & Click AUC \\
\bottomrule
\end{tabularx}
\end{table}

\myitem{Agent Environment and Computational Budget}
For each task instance $(\mathcal{I}, \pi_0, \mathcal{D}, B)$, the agent operates in a sandboxed one-GPU environment containing the task description $\mathcal{I}$, the provided recommendation baseline $\pi_0$, and the available data in $\mathcal{D}$. As shown in Figure~\ref{fig:agent_workflow}, the agent is equipped with a standard coding-agent interface consisting of local file inspection and editing, shell command execution, dependency installation, and the launching of training and validation jobs.
All operations are strictly confined to the local workspace: while the agent is free to modify the baseline strategy $\pi_0$, it is prohibited from accessing the open Internet, web browsers, or external search tools. To prevent data leakage, both the ground-truth test labels and any test-set evaluation feedback remain inaccessible throughout the run.

Notably, AutoRSE allocates a fixed wall-clock budget of $B = 12$ hours to each run. To guarantee consistent budget utilization, if an agent terminates prematurely, we force it to continue exploring further optimizations within the original constraints using a predefined continuation prompt (Appendix Section~\ref{sec:evaluation_protocol}). Upon budget expiration, the run terminates, and the final recommendation strategy $\pi_{final}$ alongside recommendation result $\hat{Y}_{final}$ are saved for evaluation. As shown in Section~\ref{sec:rq1}, this 12-hour budget is sufficient for meaningful iterative optimization, with performance trajectories largely converging by the end of the runs.

\myitem{Evaluation Protocol}
\label{sec:evaluation_protocol}
After the agent completes the task, AutoRSE invokes a task-specific verifier validates the submission format of $\hat Y_{final}$ and evaluates the predictions against the hidden test labels $Y_{\mathrm{test}}$ using the task's primary metric $\mathcal{M}$ (listed in Table~\ref{tab:tasks}). Let $S_{\mathrm{agent}} = \mathcal{M}(\hat Y_{final}, Y_{\mathrm{test}})$ and $S_{\mathrm{base}} = \mathcal{M}(Y_0, Y_{\mathrm{test}})$ denote the test scores of the agent's final submission $\pi_{final}$ and the initial baseline $\pi_0$, respectively, where $Y_0$ represents the baseline predictions under the same evaluation protocol. To quantify performance gains, we report both the absolute test score $S_{\mathrm{agent}}$ and the relative improvement ($\Delta_{\mathrm{rel}}$) over the initial baseline:
\begin{equation}
    \Delta_{\mathrm{rel}}
    =
    \frac{S_{\mathrm{agent}} - S_{\mathrm{base}}}
         {S_{\mathrm{base}}}
    \times 100\%.
\end{equation}

\subsection{Trajectory-level Analysis}
\label{sec:trajectory-analysis}

While outcome-level evaluation ($\Delta_{\mathrm{rel}}$) measures whether an RSE-Agent improves a recommendation baseline, it does not reveal how such improvements are achieved. To address this, AutoRSE incorporates a trajectory-level analysis pipeline that segments continuous agent trajectories, which weave together a sequence of code improvements and exploratory attempts, into individual experimental iterations (e.g., single parameter sweeps or architecture modifications within a run), identifies their underlying optimization strategies, and evaluates the algorithmic novelty of the final submitted algorithms.

\myitem{Experiment Segmentation}
AutoRSE records the complete agentic trajectory of each run, including agent reasoning, responses, and tool calls. To analyze the agent's iterative exploration, we first segment each continuous trajectory into a sequence of individual experiments by locating tool calls that launch model training or evaluation (\eg \texttt{python train.py}). Furthermore, we clean each segmented trajectory by filtering out uninformative operations (\eg file reads or failed tool calls) that would otherwise introduce noise and interfere with subsequent strategy identification.

\myitem{Optimization Strategy Identification}
We then employ an LLM judge (with prompts detailed in Appendix section~\ref{sec:trajectory-analysis}) to identify the optimization strategy of each segmented experiment. Through a preliminary inspection of the agent trajectories (detailed in Appendix Section ~\ref{sec:trajectory-analysis}), we observe that all experimental modifications can be systematically mapped to one of six optimization strategies, where Categories 1--3 correspond to \textbf{engineering optimizations} and Categories 4--6 represent \textbf{algorithmic innovations}:
\begin{enumerate}[leftmargin=*]
    \item[\textbf{C1.}] \textbf{Parameter Tuning:} Numerical adjustments to existing hyperparameters (\eg learning rate, embedding dimension, dropout rate, or ensemble weights) without introducing new mechanisms.
    \item[\textbf{C2.}] \textbf{Model Ensemble:} Combining multiple checkpoints or model variants via prediction averaging, weighted blending, or reranking.
    \item[\textbf{C3.}] \textbf{Training Strategy:} Adjusting the training procedure, such as swapping optimizers, altering learning-rate schedules, or applying standard regularization techniques (\eg weight decay, gradient clipping).
    \item[\textbf{C4.}] \textbf{Feature Modeling:} Designing new input representations, domain-specific feature transformations, cross-feature signals, or user behavioral sequence formulations.
    \item[\textbf{C5.}] \textbf{Model Architecture:} Structural improvements to neural network components, such as attention mechanisms, feature interaction layers, gating modules, or representation spaces.
    \item[\textbf{C6.}] \textbf{Objective Function:} Reformulating loss functions, auxiliary learning tasks, negative sampling strategies, or training targets.
\end{enumerate}

\myitem{Algorithmic Novelty Assessment}
While step-by-step strategy identification tracks the agent's iteration process, we also evaluate the overall strategy novelty of the final submitted strategy $\pi_{\text{final}}$ relative to the initial baseline $\pi_0$ to determine whether the agent's improvements represent genuine substantive innovations rather than merely superficial or repetitive adjustments.Following~\cite{bhushan2026llmagentsdiscover}, we rate the code modifications from $\pi_0$ to $\pi_{final}$ on a 0--4 ordinal scale. Specifically, \textbf{Level 0 (Routine)} denotes trivial hyperparameter tuning without structural changes; \textbf{Level 1 (Combinational)} reconfigures existing modules within $\pi_0$; \textbf{Level 2 (Exploratory)} introduces standard off-the-shelf components (\eg optimizers, schedules, or ensembles); \textbf{Level 3 (Hybrid)} integrates existing modules with newly designed mechanisms; and \textbf{Level 4 (Transformational)} represents fundamental innovations in model architecture, feature modeling, or objective functions. For each run, we employ three LLM judges to evaluate the final submitted strategy and average their ratings to obtain the final novelty score (detailed rubrics and prompts are provided in Appendix Section~\ref{app:novelty_prompt}).

\definecolor{gaincolor}{RGB}{25,120,65}
\definecolor{losscolor}{RGB}{190,45,45}
\begin{table*}[!t]
\centering
\caption{\textbf{Final performance across six recommendation tasks.}
\textnormal{Scores are reported as mean $\pm$ standard deviation
over three trials.
Bold and underlined values mark the best and second-best
results per column. Green/red arrows indicate relative
gains/drops from the corresponding baseline.
Avg.\ Gain is the unweighted mean relative change
across the six tasks.}}
\label{tab:main_results}

\newcommand{\gaincell}[3]{%
  \shortstack{$#1 \pm #2$\\[-1pt]%
  {\scriptsize\textcolor{gaincolor}{$\uparrow\,#3\%$}}}%
}
\newcommand{\losscell}[3]{%
  \shortstack{$#1 \pm #2$\\[-1pt]%
  {\scriptsize\textcolor{losscolor}{$\downarrow\,#3\%$}}}%
}

\setlength{\tabcolsep}{4pt}
\renewcommand{\arraystretch}{1.25}
\resizebox{\textwidth}{!}{%
\begin{tabular}{@{}lccccccc@{}}
\toprule
\multirow{3}{*}{LLM $\times$ Harness}
& \multicolumn{3}{c}{Generative Recommendation}
& \multicolumn{3}{c}{CTR Prediction}
& \multirow{3}{*}{Avg.\ Gain } \\
\cmidrule(lr){2-4}\cmidrule(lr){5-7}
& HSTU & IntSR & TIGER
& OneTrans & RankMixer & WuKong & \\
& NDCG@10 
& Macro NDCG@10 
& NDCG@10 
& AUC 
& AUC 
& AUC  & \\
\midrule

Baseline
& 0.0535 & 0.6940 & 0.0332
& 0.6767 & 0.7859 & 0.9645
& 0.00\% \\
\midrule

Claude Code $\times$ DeepSeek V4 Flash
& \gaincell{0.0674}{0.0032}{25.98}
& \losscell{0.6606}{0.0541}{4.81}
& \gaincell{0.0476}{0.0138}{43.37}
& \gaincell{\mathbf{0.7467}}{0.0324}{10.34}
& \gaincell{0.7979}{0.0015}{1.53}
& \gaincell{0.9805}{0.0030}{1.66}
& \textcolor{gaincolor}{+13.01\%} \\[3pt]

Claude Code $\times$ Claude Opus 4.8
& \gaincell{\mathbf{0.0754}}{0.0043}{40.93}
& \gaincell{\underline{0.7325}}{0.0292}{5.55}
& \gaincell{\underline{0.0631}}{0.0164}{90.06}
& \gaincell{0.7184}{0.0068}{6.16}
& \gaincell{\mathbf{0.8002}}{0.0009}{1.82}
& \gaincell{\underline{0.9840}}{0.0039}{2.02}
& \textcolor{gaincolor}{\textbf{+24.42\%}} \\[3pt]

Codex $\times$ DeepSeek V4 Flash
& \gaincell{0.0657}{0.0030}{22.80}
& \losscell{0.6485}{0.0335}{6.56}
& \gaincell{0.0583}{0.0147}{75.60}
& \gaincell{0.7203}{0.0044}{6.44}
& \gaincell{0.7974}{0.0003}{1.46}
& \gaincell{0.9790}{0.0018}{1.50}
& \textcolor{gaincolor}{+16.88\%} \\[3pt]

Codex $\times$ GPT-5.5
& \gaincell{0.0672}{0.0065}{25.61}
& \gaincell{\mathbf{0.7692}}{0.1904}{10.84}
& \gaincell{\mathbf{0.0659}}{0.0046}{98.49}
& \gaincell{\underline{0.7292}}{0.0342}{7.76}
& \gaincell{0.7954}{0.0030}{1.21}
& \gaincell{0.9796}{0.0032}{1.57}
& \textcolor{gaincolor}{\underline{+24.24\%}} \\[3pt]

Terminus-2 $\times$ DeepSeek V4 Flash
& \gaincell{0.0667}{0.0057}{24.67}
& \losscell{0.6291}{0.0858}{9.35}
& \gaincell{0.0554}{0.0158}{66.87}
& \losscell{0.6665}{0.0447}{1.51}
& \gaincell{0.7977}{0.0012}{1.50}
& \gaincell{\mathbf{0.9877}}{0.0020}{2.41}
& \textcolor{gaincolor}{+14.10\%} \\[3pt]

Terminus-2 $\times$ GPT-5.5
& \gaincell{\underline{0.0697}}{0.0058}{30.28}
& \losscell{0.6146}{0.0711}{11.44}
& \gaincell{0.0614}{0.0053}{84.94}
& \gaincell{0.7285}{0.0498}{7.65}
& \gaincell{0.7944}{0.0021}{1.08}
& \gaincell{0.9749}{0.0035}{1.08}
& \textcolor{gaincolor}{+18.93\%} \\[3pt]

Terminus-2 $\times$ Claude Opus 4.8
& \gaincell{0.0636}{0.0015}{18.88}
& \losscell{0.6761}{0.0334}{2.58}
& \losscell{0.0273}{0.0190}{17.77}
& \gaincell{0.7207}{0.0232}{6.50}
& \gaincell{\underline{0.7989}}{0.0014}{1.65}
& \gaincell{0.9823}{0.0004}{1.85}
& \textcolor{gaincolor}{+1.42\%} \\

\bottomrule
\end{tabular}%
}
\end{table*}
\section{Experiments}

In this section, we conduct experiments to answer the following research questions:
\begin{itemize}
    \item \textbf{RQ1:} Can LLM agents autonomously improve performance across diverse recommendation scenarios?
    \item \textbf{RQ2:} If so, how do current RSE-Agents achieve such improvements?
    \item \textbf{RQ3:} How can we further improve the performance of current RSE-Agents?
\end{itemize}

\subsection{Experimental Setup}
\label{sec:exp_setup}

\subsubsection{Evaluated RSE-Agents}
We evaluate three frontier LLMs, namely Claude Opus 4.8, GPT-5.5, and DeepSeek V4 Flash, paired with three widely used agent harnesses, which equip the LLMs with execution infrastructure for RSE, including \texttt{Claude Code}, \texttt{Codex}, and \texttt{Terminus-2}. Since \texttt{Claude Code} does not support the OpenAI API endpoint and \texttt{Codex} does not support the Anthropic API endpoint, \texttt{Claude Code $\times$ GPT-5.5} and \texttt{Codex $\times$ Claude Opus 4.8} are excluded. We finally obtain seven LLM--harness combinations.

\subsubsection{Implementation Details}  
To ensure reliable evaluation, we conduct three independent runs for each task--agent combination. Every run is executed in an isolated container equipped with a single NVIDIA H200 GPU (141\,GB HBM3e), 16 vCPUs, and 32\,GB RAM under a strict 12-hour wall-clock budget. For the agent harnesses, we use \texttt{Claude Code} v2.1.226, \texttt{Codex} (\texttt{codex-cli} v0.147.0), and \texttt{Terminus-2} v0.20.0.
For the trajectory-level analysis, we employ DeepSeek V4 Flash as the LLM judge for experiment segmentation, strategy identification, and algorithmic novelty scoring.

\subsection{Performance Improvement over Reference Baselines (RQ1)}
\label{sec:rq1}

\subsubsection{Main Results}
To evaluate whether LLM agents can autonomously improve recommendation algorithms, we first benchmark three frontier LLMs (Claude Opus 4.8, GPT-5.5, and DeepSeek V4 Flash) paired with three widely used agent harnesses (Claude Code, Codex, and Terminus-2) across the six tasks in AutoRSE. For each setting, we conduct three independent trials and report the mean and standard deviation of the final test performance. The results are shown in Table~\ref{tab:main_results}, from which we draw the following observations: 
(1) \textbf{RSE-Agents consistently improve recommendation performance, confirming their practical viability.} Out of the 42 evaluated settings, agents outperform the reference baselines in 35 cases. Specifically, on five of the six tasks (HSTU, TIGER, OneTrans, RankMixer, and WuKong), almost all LLM--harness combinations yield positive gains over strong industrial-grade baselines, achieving a mean relative improvement of 16.1\% across all tasks and configurations. These results suggest that current LLM agents can already serve as effective autonomous engineers for improving existing recommender systems. However, agents primarily explore engineering-level
optimizations, with limited algorithmic innovation(as detailed in \S\ref{sec:rq2}); 
(2) \textbf{Agent performance heavily depends on both the backbone LLM and the harness.} Across all tasks, frontier proprietary models paired with
their native harnesses---\texttt{Claude Code $\times$
Claude Opus 4.8} (+24.42\%) and \texttt{Codex $\times$
GPT-5.5} (+24.24\%)---deliver the strongest and most reliable
improvements, ranking first or second on nearly every task.
However, when the same models are deployed within a
third-party harness (\texttt{Terminus-2}), their performance
declines substantially, particularly for Claude Opus 4.8.
These results suggest that both the backbone LLM and the
harness influence agents' optimization performance; and
(3) \textbf {Performance improvements vary substantially across recommendation scenarios and task.} RSE-Agents achieve an average gain of 29.2\% on generative recommendation tasks compared to 3.1\% on CTR prediction. We attribute this discrepancy to the differing maturity and metric saturation of the two recommendation scenarios: CTR prediction is a well-established task bounded by the AUC metric (\eg WuKong's baseline is already 0.9645, leaving limited headroom), whereas generative recommendation is an emerging paradigm where baseline training recipes and hyperparameters are far from saturated, offering vast room for optimization. Conversely, on the joint search--recommendation task IntSR, five of the seven agents underperform the baseline. Because IntSR involves a complex multi-task optimization landscape, generic tuning strategies easily lead to negative transfer or validation overfitting, which motivates a deeper analysis of how agents actually optimize models.

\subsubsection{Optimization Dynamics over Time}
\begin{figure}[!t]
    \centering
    \includegraphics[width=\columnwidth]{
        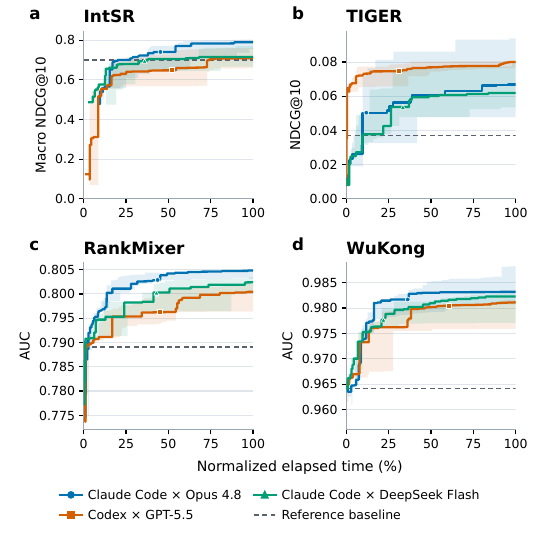
    }
    \caption{Progression of best-so-far validation performance over normalized elapsed time (12 hours = 100\%). Solid lines and shaded regions denote the median and min--max range across three independent runs, respectively; dashed lines indicate the initial reference baselines.}
    \label{fig:iteration_curves}
\end{figure}

To examine how agent performance evolves during the optimization process, we track the best-so-far validation score over the 12-hour run (with elapsed time normalized from 0\% to 100\%) for three representative agents: \texttt{Claude Code $\times$ Claude Opus 4.8}, \texttt{Codex $\times$ GPT-5.5}, and \texttt{Claude Code $\times$ DeepSeek V4 Flash}. Figure~\ref{fig:iteration_curves} plots the median validation performance across three runs (solid lines) and the min--max range (shaded areas) on four representative tasks (complete results for all six tasks are provided in Appendix~\ref{fig:appendix_trajectories}), from which we observe that:
(1) \textbf{Performance improves rapidly in the early stages before entering a plateau, confirming that the 12-hour budget is sufficient.} Across the tasks, agents quickly surpass the reference baseline (dashed gray line) within the first 10\%--25\% of the time budget and achieve the vast majority of their gains before the 50\% mark (6 hours). In the second half of the runs, the trajectories largely flatten out with diminishing marginal returns, empirically validating our 12-hour budget design in Section~\ref{sec:framework}; 
(2) \textbf{Repeated runs of the same agent exhibit substantial variability throughout the optimization process.} As shown by the wide shaded bands on tasks such as TIGER and WuKong (Figures~\ref{fig:iteration_curves}b and \ref{fig:iteration_curves}d), the performance gap between different runs emerges early and persists until the end of the 12-hour budget, converging to noticeably different plateaus. This suggests that agents may get stuck refining the same idea,
making repeated small changes with limited gains. Different runs may therefore plateau at very different scores; and
(3) \textbf{On complex tasks such as IntSR, weaker agents struggle to surpass the reference baseline even on the validation set.}All three agents quickly surpass the baseline on TIGER,
RankMixer, and WuKong (Figure~\ref{fig:iteration_curves}b--d),
but diverge on IntSR (Figure~\ref{fig:iteration_curves}a):
only \texttt{Claude Code $\times$ Claude Opus 4.8} achieves
substantial gains, while the other two remain near the
baseline. This contrast accompanies the mixed test results
on IntSR in Table~\ref{tab:main_results}.
Inspection of unsuccessful runs on IntSR, which uses a unified
model for both search and recommendation, suggests that agents
struggle to balance the two tasks, as improving accuracy on one
may reduce accuracy on the other and lead to poor overall
performance.

\subsection{Trajectory-Level Analysis of Agent Improvements (RQ2)}
\label{sec:rq2}
\subsubsection{Distribution of Optimization Strategies.} 
To understand how RSE-Agents achieve the performance gains observed in RQ1, we apply our trajectory-level analysis pipeline (\S\ref{sec:trajectory-analysis}) to categorize a total of 4425 segmented experiments across all agent runs into the six optimization strategies. The overall distribution is presented in Figure~\ref{fig:improvement_types}, with the detailed breakdown for each LLM-harness combination provided in Appendix Table~\ref{tab:strategy_distribution}. From these results, we draw two key observations: 
(1) \textbf{Current RSE-Agents overwhelmingly rely on surface-level engineering optimizations rather than substantive algorithmic innovations.} As shown in Figure~\ref{fig:improvement_types}, engineering optimizations account for \textbf{91.7\%} of all segmented experiments—dominated by Model Ensemble (59.0\%) and Parameter Tuning (26.2\%), followed by Training Strategy (6.6\%). In contrast, genuine algorithmic innovations represent only \textbf{8.3\%} of the experiments, with Feature Modeling, Model Architecture, and Objective Function accounting for merely 4.2\%, 3.1\%, and 1.0\%, respectively; and 
(2) \textbf{Although different agents prefer different engineering tricks, all of them rarely explore algorithmic innovations.} When breaking down the strategies by each LLM--harness combination (Appendix Table~\ref{tab:strategy_distribution}), we observe clear differences in which engineering strategy an agent prefers. For example, with GPT-5.5, \texttt{Codex} mainly focuses on Parameter Tuning, whereas \texttt{Terminus-2} mainly relies on Model Ensemble. Similarly, with Claude Opus 4.8, \texttt{Claude Code} spends 83.2\% of its experiments on Model Ensemble, compared to 53.2\% under \texttt{Terminus-2}. However, regardless of whether an agent prefers parameter tuning or ensembling, every single LLM--harness combination spends over 80\% of its experiments on engineering optimizations, confirming that the lack of algorithmic innovation is a universal bottleneck across all tested agents.

\begin{figure}[t]
    \centering
    \includegraphics[width=0.75\columnwidth]{
        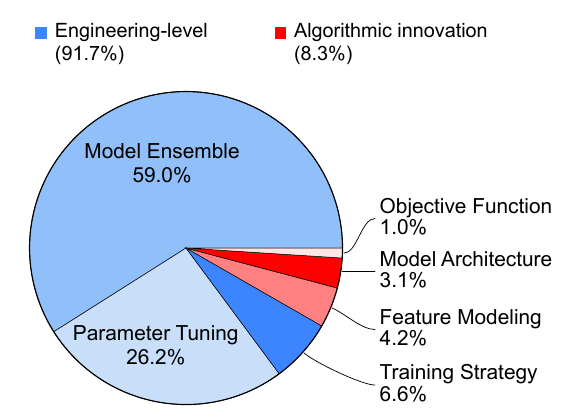
    }
    \caption{Distribution of segmented experiments
    across the six optimization strategies.}
    \label{fig:improvement_types}
\end{figure}

\subsubsection{Case Study: Algorithmic Innovation vs. Hyperparameter Tuning}

To examine whether algorithmic modifications can yield performance gains beyond hyperparameter optimization, we analyze two representative agent trajectories from the HSTU run (\texttt{Claude Code × Claude Opus 4.8}). Both variants are benchmarked against a shared control baseline under identical training conditions.

\begin{casestudy}{Case 1: Parameter Tuning \& Training Strategy on HSTU}
Starting from a baseline validation NDCG@10 of 0.06770 on HSTU, the agent conducts targeted experiments across hyperparameters and training strategies while preserving the architecture. While aggressive regularization yields mixed results, introducing a cosine learning-rate schedule with warmup achieves a peak NDCG@10 of 0.06997 (+3.4\% relative gain). This trajectory highlights both the efficacy of systematic parameter tuning and its inherent performance ceiling, motivating feature-level innovations.
\end{casestudy}

\begin{casestudy}{Case 2: Feature Modeling via Content Fusion in HSTU}
To explore gains beyond parameter and strategy tuning, the agent evaluates a feature modeling variant on the same reference HSTU model. By introducing a learnable projection that maps frozen 768-dimensional item features into the ID embedding space, the agent incorporates content-informed representations into both sequence encoding and candidate scoring while retaining the core HSTU backbone. Under identical reference training settings, this feature modeling approach achieves a validation NDCG@10 of 0.07216—a ~6.6\% improvement over the baseline and a 3.1\% gain over the best variant discovered in Case~1.
\end{casestudy}

This comparison shows that while \textbf{Training Strategy} and \textbf{Parameter Tuning} can improve performance by adjusting external configurations, they do not address the structural limitations of the model itself. By enhancing internal representations through \textbf{Model Architecture} innovations, the agent achieves substantial additional gains, highlighting the vast potential of algorithmic exploration.

\begin{figure}[t]
    \centering
    \includegraphics[width=\columnwidth]{
        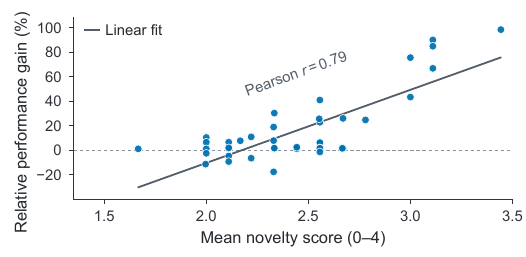
    }
    \caption{Relationship between the mean algorithmic novelty score (0--4 scale) and mean relative test improvement ($\Delta_{\mathrm{rel}}$) across all evaluated task--agent combinations.}
    \label{fig:novelty_results}
\end{figure}

\subsubsection{Algorithmic Novelty vs.\ Performance Gains}
To further verify whether algorithmic innovations lead to larger performance gains across all tasks, we score the algorithmic novelty (on the 0--4 scale defined in \S\ref{sec:trajectory-analysis}) of each run's final submitted algorithm and examine its relationship with the relative test improvement ($\Delta_{\mathrm{rel}}$). For each task and LLM--harness combination, Figure~\ref{fig:novelty_results} plots the mean novelty score against the mean $\Delta_{\mathrm{rel}}$, from which we observe that: 
(1) \textbf{Current RSE-Agents achieve a modest average novelty score of 2.42 out of 4.0, confirming their limited algorithmic innovation.} Consistent with our strategy distribution findings, final submitted algorithms fall around Level~2 (\emph{Exploratory}), meaning that agents mainly combine standard off-the-shelf components (\eg ensembles and learning-rate schedules) rather than designing Level~3 (\emph{Hybrid}) or Level~4 (\emph{Transformational}) algorithmic innovations; and 
(2) \textbf{Higher algorithmic novelty is strongly correlated with larger performance gains ($r = 0.79$).} As shown in Figure~\ref{fig:novelty_results}, the novelty score exhibits a strong positive Pearson correlation with $\Delta_{\mathrm{rel}}$. Solutions with higher novelty scores consistently achieve greater relative improvements over the baselines, demonstrating that algorithmic innovation is key to unlocking higher performance ceilings.


\subsection{Enhancing Agent Performance with Injected Ideas (RQ3)}

%
%

Our analysis in RQ2 reveals that current agents rarely explore high-novelty algorithmic innovations, which may limit their performance ceiling. To investigate whether explicit algorithmic guidance can unlock further gains, we conduct idea-injection experiments, comparing agents guided by external research ideas against those relying on self-generated ideas.

\subsubsection{Experimental Setup}
\begin{figure}[t]
    \centering
    \includegraphics[width=\columnwidth]
    {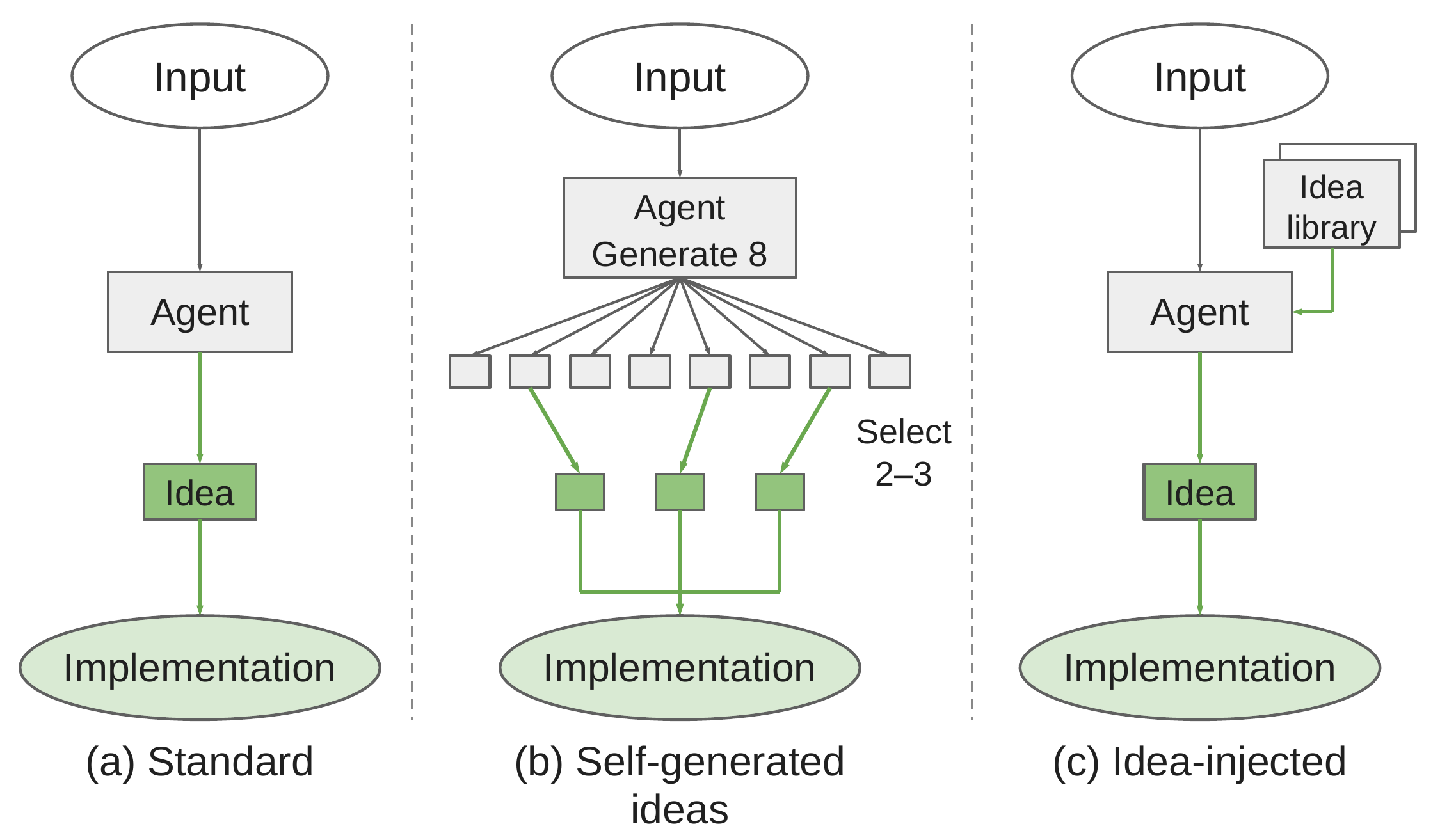}
    \caption{Overview of the three experimental settings
    for idea generation and implementation.}
    \label{fig:idea_settings}
\end{figure}

As illustrated in Figure~\ref{fig:idea_settings}, we evaluate Claude Code $\times$ DeepSeek V4 Flash under three ideation settings:
(1) \textbf{\emph{Standard}}, which directly follows the default autonomous optimization protocol in \S\ref{sec:exp_setup}; 
(2) \textbf{\emph{Self-generated ideas}}, which incorporates an explicit \emph{generate-and-select} (GenSelect~\cite{toshniwal2025genselect}) stage—in each ideation round, the agent first brainstorms eight candidate ideas and then selects two or three promising directions to pursue; and 
(3) \textbf{\emph{Idea-injected}}, which provides the agent with an external idea library as additional context at the start of each run, allowing it to autonomously select and adapt relevant research directions. Specifically, we construct this idea library from papers published at top-tier venues (\eg WWW, SIGIR, and KDD) as well as publicly shared winning solutions and code implementations from recommender system competitions. From these materials, we distill core methodological designs and practical techniques—spanning model architectures, data processing, training strategies, and inference—into concise textual entries covering both task-specific designs and general recommendation techniques.

\subsubsection{Results and Analysis.}
Table~\ref{tab:idea_injection} reports the final submission scores and novelty scores across three independent trials, from which we draw the following observations:
(1) \textbf{Expanding self-generated ideas yields minimal or even negative gains, indicating that current LLMs struggle to propose effective research ideas on their own.} As shown in Table~\ref{tab:idea_injection}, despite explicitly generating and selecting among multiple candidate ideas, \emph{Self-generated ideas} achieves only marginal improvements over \emph{Standard} on most tasks and even suffers performance degradation on IntSR ($-8.08\%$) and RankMixer ($-0.49\%$). Because the self-proposed ideas still follow similar low-novelty directions, simply generating and filtering more candidates fails to produce better solutions; 
(2) \textbf{With injected high-quality ideas, agent performance improves substantially, demonstrating that idea quality is the primary bottleneck.} Across all six tasks, \emph{Idea-injected} consistently outperforms both \emph{Standard} and \emph{Self-generated ideas}, achieving average score gains of 14.23\% and 13.35\% and novelty gains of 0.76 and 0.58 points, respectively. This indicates that code implementation is not the bottleneck for current RSE-Agents: if they could generate high-quality ideas comparable to the injected ones, they would achieve much stronger performance. Thus, strengthening autonomous ideation is a critical direction for future RSE-Agents; and 
(3) \textbf{The value of research guidance depends on how well the proposed directions match the specific weaknesses of the baseline, beyond the extent of methodological change.} By analyzing agents' research trajectories (Appendix~\ref{app:guidance_cases}), we find that they successfully implement promising directions through concrete code changes—for example, integrating content embeddings into HSTU's shared input and scoring representations, and expanding IntSR's training candidates with in-batch and popularity-based negatives. Ideas that introduce previously underused predictive information or directly address weaknesses in representation learning and optimization produce substantial gains, whereas modifications to components that are not the main bottleneck may increase solution novelty while offering only marginal score improvements.

\begin{table*}[t]
\centering
\small

\providecolor{gaincolor}{RGB}{25,120,65}
\providecolor{losscolor}{RGB}{190,45,45}

\providecommand{\ideagaincell}[2]{%
  \shortstack{$#1$\\[-1pt]%
  {\scriptsize\textcolor{gaincolor}{$\uparrow\,#2\%$}}}}

\providecommand{\idealosscell}[2]{%
  \shortstack{$#1$\\[-1pt]%
  {\scriptsize\textcolor{losscolor}{$\downarrow\,#2\%$}}}}

\renewcommand{\ideagaincell}[2]{%
  \begin{tabular}[t]{@{}c@{}}
    $#1$ \\[-3pt]
    {\scriptsize\textcolor{gaincolor}{$\uparrow\,#2\%$}}
  \end{tabular}%
}

\renewcommand{\idealosscell}[2]{%
  \begin{tabular}[t]{@{}c@{}}
    $#1$ \\[-3pt]
    {\scriptsize\textcolor{losscolor}{$\downarrow\,#2\%$}}
  \end{tabular}%
}

\caption{Final test scores and algorithmic novelty scores under the \emph{Standard}, \emph{Self-generated ideas}, and \emph{Idea-injected} settings. \textnormal{Arrows and percentages indicate relative changes compared to the \emph{Standard} setting.}}
\label{tab:idea_injection}

\setlength{\tabcolsep}{3pt}
\renewcommand{\arraystretch}{1.15}

\begin{tabular*}{\textwidth}{
    @{}l@{\extracolsep{\fill}}*{12}{c}@{}
}
\toprule
Setting
& \multicolumn{2}{c}{HSTU}
& \multicolumn{2}{c}{IntSR}
& \multicolumn{2}{c}{TIGER}
& \multicolumn{2}{c}{OneTrans}
& \multicolumn{2}{c}{RankMixer}
& \multicolumn{2}{c}{WuKong} \\

\cmidrule(lr){2-3}
\cmidrule(lr){4-5}
\cmidrule(lr){6-7}
\cmidrule(lr){8-9}
\cmidrule(lr){10-11}
\cmidrule(lr){12-13}

& Score & Novelty
& Score & Novelty
& Score & Novelty
& Score & Novelty
& Score & Novelty
& Score & Novelty \\
\midrule

Standard
& 0.06740 & 2.67
& 0.66060 & 2.11
& 0.04760 & 3.00
& 0.74670 & 2.00
& 0.79790 & 2.11
& 0.98050 & 2.56 \\
\addlinespace[3pt]

Self-generated ideas
& \ideagaincell{0.06810}{1.04} & \ideagaincell{3.00}{12.36}
& \idealosscell{0.60724}{8.08} & \idealosscell{2.00}{5.21}
& \ideagaincell{0.05190}{9.03} & \ideagaincell{3.33}{11.00}
& \ideagaincell{0.76109}{1.93} & \ideagaincell{2.33}{16.50}
& \idealosscell{0.79403}{0.49} & \ideagaincell{2.33}{10.43}
& \ideagaincell{0.98931}{0.90}
& \begin{tabular}[t]{@{}c@{}}
    $2.56$ \\[-3pt]
    {\scriptsize\textcolor{gray}{$0.00\%$}}
  \end{tabular} \\

Idea-injected
& \ideagaincell{0.08310}{23.29} & \ideagaincell{3.33}{24.84}
& \ideagaincell{0.75707}{14.60} & \ideagaincell{3.00}{42.18}
& \ideagaincell{0.06729}{41.36} & \ideagaincell{3.67}{22.33}
& \ideagaincell{0.77729}{4.10} & \ideagaincell{2.67}{33.33}
& \ideagaincell{0.79992}{0.25} & \ideagaincell{3.00}{42.18}
& \ideagaincell{0.99786}{1.77} & \ideagaincell{3.33}{30.38} \\

\bottomrule
\end{tabular*}
\end{table*}
\section{Related Work}

\subsection{Recommender System Automation}


Automating recommender system development has long been studied through
automated machine learning and neural architecture search.
AutoRec~\citep{wang2020autorec} automates model search hyperparameter
tuning for recommendation tasks.
AutoFIS~\citep{liu2020autofis} automatically selects informative feature
interactions in factorization models, while
AutoCTR~\citep{song2020autoctr} searches for interaction architectures
using evolutionary optimization.
These approaches reduce manual effort by optimizing model configurations
or components within predefined search spaces, but leave research
hypothesis formulation and the broader experimental workflow largely
to human engineers.

With advances in LLM-based agents, recent work has extended recommendation
automation to iterative research workflows. for example, 
AgentX\citep{lao2026agentx} coordinates specialized agents to generate
improvement proposals, implement code changes, evaluate online
experiments, and accumulate experimental knowledge.
Self-Evolving Recommendation
System~\citep{wang2026selfevolving} combines an offline agent for
hypothesis generation and proxy-metric optimization with an online agent
that validates candidates against delayed business metrics.
Its optimization targets include model architectures, optimizers,
and reward functions.
These pioneering systems demonstrate the practical feasibility of
agent-driven recommendation iteration within their respective
proprietary industrial pipelines.
However, their system-specific evaluations leave open how different
RSE-Agents compare under a shared protocol and which improvement
strategies account for their performance gains across tasks.

Our work addresses these questions through AutoRSE, a unified
benchmark spanning six recommendation tasks with standardized execution
environments, resource budgets, and hidden-test evaluation.
Beyond final performance, we analyze complete agent trajectories to
characterize their proposed changes and relate improvement strategies
and research effort to experimental outcomes.
This analysis reveals that current agents predominantly rely on
model ensemble and parameter tuning, while changes to
architectures and objectives function are comparatively rare.
Motivated by this finding, we introduce a idea-injection experiment and selection module to support the exploration of
substantive algorithmic improvements before costly implementation
and training.

\subsection{LLM Agents for Engineering Tasks}


LLM-based agents have increasingly been applied to software
engineering (SWE) and machine learning engineering (MLE).
SWE-bench~\cite{jimenez2024swebench} evaluates models on
real-world GitHub issues, while SWE-agent~\cite{yang2024sweagent}
enables autonomous repository navigation, code editing, and
test execution through a dedicated agent--computer interface.
Moving beyond software maintenance,
MLAgentBench~\cite{huang2024mlagentbench} evaluates agents
on iterative machine learning experimentation, and
MLE-bench~\cite{chan2024mlebench} measures their ability to
solve competition-derived tasks involving data preparation,
model training, and performance optimization.
Recent work further extends these capabilities toward
autonomous scientific research.
The AI Scientist~\cite{lu2024aiscientist} connects idea
generation, experiment execution, and manuscript writing,
while The AI Scientist-v2~\cite{yamada2025aiscientistv2}
uses agentic tree search to organize and refine experiments.
Agent Laboratory~\cite{schmidgall2025agentlaboratory}
coordinates literature review, experimentation, and report
writing around human-proposed research ideas.
AI-Researcher~\cite{tang2025airesearcher} integrates literature
analysis, algorithm implementation, and experimental validation,
and introduces Scientist-Bench for guided and open-ended
research tasks.

Within this broader research agenda, recommender systems
engineering (RSE) provides a domain-specific setting for
evaluating agents' ability to improve algorithms.
Our RSE setting requires sustained optimization of faithfully
reproduced, strong recommendation baselines under a fixed
computational budget. Agents must determine what to change,
formulate useful hypotheses, and evaluate their implementations
through experimentation. Costly training and noisy validation
feedback make unsuccessful directions expensive and require
agents to coordinate hypothesis generation, implementation,
and evaluation over long horizons. This setting captures
practical challenges of recommendation research and allows
us to examine whether agents can produce substantive
algorithmic advances beyond routine tuning and engineering
heuristics. Beyond final performance, our trajectory-level
analysis examines the modifications agents implement,
providing a basis for understanding their research strategies
and the limitations of their algorithmic exploration.
\section{Conclusion}

In this work, we introduce AutoRSE, a unified framework for evaluating
RSE-Agents. Using this framework, we conduct a
large-scale evaluation across diverse LLM-Agents,
and recommendation tasks. Our experiments show that
LLM agents can autonomously improve recommendation
performance. And further trajectory analysis reveals that agents
primarily explore model ensembling and parameter tuning,
while changes to architectures and learning objectives
remain uncommon. These findings highlight a lack of substantive algorithmic
innovation in the solutions developed by current RSE-Agents. Motivated by the limited
exploration of algorithmic ideas, we further conduct idea-injection experiments.
Our idea-injection experiments further show that ideas
from frontier research and human experts yield larger
performance gains than agents' self-generated ideas,
highlighting idea generation as a key bottleneck and
motivating stronger algorithmic ideation capabilities
in future RSE-Agents.

In the future, we will continuously expand AutoRSE with more
diverse recommendation tasks, larger-scale evaluation protocols,
and a growing, community-driven leaderboard to foster open and
reproducible assessment of RSE-Agents. Additionally, we aim to
investigate advanced reasoning and reflection mechanisms to en-
rich the creativity and quality of agents’ self-generated hypotheses,
bridging the gap between automated software engineering and
substantive algorithmic breakthroughs. We hope AutoRSE serves
as a solid and reliable stepping stone toward achieving fully autonomous AI scientists in future web recommendation research and application.

\bibliographystyle{ACM-Reference-Format}
\bibliography{main}

\clearpage
\appendix
\tcbset{
  autorseprompt/.style args={#1}{
    reset,
    nofloat,
    visible,
    natural height,
    enhanced jigsaw,
    breakable,
    width=\linewidth,
    colback=black!2,
    colframe=black!55,
    colbacktitle=black!72,
    coltitle=white,
    coltext=black,
    boxrule=0.4pt,
    titlerule=0pt,
    arc=1mm,
    outer arc=1mm,
    left=6pt,
    right=6pt,
    top=5pt,
    bottom=6pt,
    toptitle=4pt,
    bottomtitle=4pt,
    title={#1},
    title after break={#1\ \textnormal{(continued)}},
    fonttitle=\normalfont\bfseries\small,
    fontupper=\normalfont\small,
    before upper={
      \setlength{\parindent}{0pt}
      \setlength{\parskip}{3pt}
      \setlength{\emergencystretch}{2em}
    },
    before skip=8pt,
    after skip=8pt
  }
}

\providecommand{\autorserecordrole}[1]{}
\renewcommand{\autorserecordrole}[1]{%
  \par\addvspace{3pt}%
  \noindent\textbf{#1}\par\nobreak
}

\providecommand{\autorsevar}[1]{}
\renewcommand{\autorsevar}[1]{%
  \texttt{\{\detokenize{#1}\}}%
}

\raggedbottom

\section{Prompt Templates}
\label{app:prompts}

The prompt templates are provided below.

\subsection{Task Launch Prompt}
\label{app:task_prompts}

The task launch prompt includes task instructions and execution
requirements.

\begin{tcolorbox}[autorseprompt={Task Launch Prompt}]

\autorserecordrole{Task}

Improve the editable \autorsevar{TASK} baseline on
\autorsevar{DATASET}.\par

\autorsevar{BASELINE_DESCRIPTION}\par

\autorserecordrole{Fixed Protocol}

\noindent\textbullet\
Use the supplied inputs in \autorsevar{INPUT_DIR}.\par

\noindent\textbullet\
Follow \autorsevar{DATA_PROTOCOL} and
\autorsevar{EVALUATION_PROTOCOL}.\par

\noindent\textbullet\
Select experimental configurations using
\autorsevar{VALIDATION_METRIC} on the public validation split.\par

\noindent\textbullet\
The reward is hidden-test \autorsevar{TEST_METRIC};
\autorsevar{DIAGNOSTIC_METRICS} are reported.\par

\autorserecordrole{Required Output}

Create \autorsevar{SUBMISSION_PATH} following
\autorsevar{SUBMISSION_SCHEMA}.
Preserve the public test sample ordering and all
task-specific validity constraints.\par

\autorserecordrole{Implementation Constraints}

Start with \autorsevar{BASELINE_README}.
You may edit the model and training pipeline, but must
not access private labels or modify the public test inputs.\par

\autorserecordrole{Execution Requirements}

PERSISTENT EXECUTION CONTRACT:\par

\noindent\textbullet\
The research phase lasts at least
\autorsevar{RESEARCH_SECONDS} seconds.\par

\noindent\textbullet\
A valid submission or an apparent improvement does not
end the run.\par

\noindent\textbullet\
Continue designing, running, and interpreting useful
experiments.\par

\noindent\textbullet\
Keep the best validation-selected submission updated
after every successful experiment so a later timeout
cannot lose it.\par

\noindent\textbullet\
The research deadline is hard: the supervisor may
interrupt the current action when it arrives, then
starts a separate finalization phase of at most
\autorsevar{FINALIZATION_SECONDS} seconds.\par

\noindent\textbullet\
During the last 30 minutes, do not start work expected
to cross the research deadline and do not use blocking
waits longer than 60 seconds. Make sure a valid
best-so-far submission already exists.\par

\noindent\textbullet\
If you return early, the supervisor will resume this
same session.\par

\noindent\textbullet\
This trial's fixed research seed is \autorsevar{SEED}.
Use it for Python, NumPy, PyTorch, data loaders, and
every comparable training run; do not switch seeds
within this trial.\par

\noindent\textbullet\
Use only the GPUs assigned and visible in this trial.
First verify the device count with nvidia-smi and do
not launch work outside the assigned GPUs.\par

\end{tcolorbox}

\subsection{Execution Control Prompts}
\label{app:execution_prompts}

The supervisor sends a continuation prompt if an agent
returns before the research deadline and a finalization
prompt when the research phase ends.

\begin{tcolorbox}[autorseprompt={Prompt: Research Continuation}]

\autorserecordrole{Instruction}

The persistent research time budget is not exhausted.
Approximately \autorsevar{REMAINING_SECONDS} seconds remain.
Resume the same research process now. Inspect the existing
experiment history, derive the next evidence-driven
modification, run it, compare validation results, and
preserve the best submission. Do not finish or merely
summarize while useful experiments remain.\par

\end{tcolorbox}

\subsection{Idea Generation and Selection}
\label{app:genselect_prompts}

\begin{tcolorbox}[autorseprompt={Prompt: Idea Generation and Selection}]

\autorserecordrole{Inputs and Evidence}

Use only task instructions, current source code,
supplied public train/\allowbreak{}validation inputs,
baseline public validation measurements, and this
trial's experiment memory. Do not retrieve papers,
search the web, access external knowledge bases,
load experience cards, title catalogs, historical
distilled memories, or invoke
retrieval/\allowbreak{}ideation Skills.
Hidden test labels and verifier feedback are forbidden.
Each round must contain exactly eight fresh ideas;
freeze two or three selections before implementation.
Record each result separately and close the round
before proposing the next batch. All selected
experiments use the original seed, model, resource
and public-validation contract.
No unrelated data or model substitution.\par

\autorserecordrole{Planning Instructions}

ROUND \autorsevar{ROUND}: PLANNING ONLY.
Do not edit implementation or launch experiments.
Write \autorsevar{PLAN_PATH} with exactly 8 ideas,
unique ids, hypothesis, evidence,
validation\_\allowbreak{}plan,
selection\_\allowbreak{}reason for every idea,
selected\_\allowbreak{}ids containing exactly 2 or 3 IDs,
results=[].
Use prior rounds as local evidence.
Return immediately after saving the plan.
The supervisor validates and freezes this plan before
any execution.\par

\end{tcolorbox}

\begin{tcolorbox}[autorseprompt={Prompt: Selected-Experiment Execution}]

\autorserecordrole{Execution and Reporting}

Execute ONLY selected idea \autorsevar{IDEA_ID}
from this frozen plan:\par

\autorsevar{FROZEN_PLAN}\par

Use seed \autorsevar{SEED}.
Save rollback identity before changing code.
Run the original public validation, keep the best valid
submission, stop your experiment's background processes,
and plan this selected item within the trial's remaining time.
Record unfinished work honestly when time runs out;
do not call a timeout a successful experiment.
Save one result JSON to \autorsevar{RESULT_PATH}.
Required:
idea\_\allowbreak{}id,
status (completed or invalid),
command,
files\_\allowbreak{}changed,
validation\_\allowbreak{}before,
validation\_\allowbreak{}after,
diagnosis,
decision.
Do not execute other ideas or start the next round.
Return after this one result is written.\par

\end{tcolorbox}

\subsection{Solution Novelty Scoring Prompt}
\label{app:novelty_prompt}

This template is reconstructed from the novelty-scoring protocol.

\begin{tcolorbox}[autorseprompt={Solution Novelty Scoring Prompt}]

\autorserecordrole{Task}

Assess how substantially the final submitted solution
departs from the original baseline. Evaluate only changes
used to produce the final predictions.\par

\autorserecordrole{Inputs}

\noindent\textbullet\
Task description: \autorsevar{TASK_DESCRIPTION}\par

\noindent\textbullet\
Baseline-to-final code diff:
\autorsevar{FINAL_CODE_DIFF}\par

\noindent\textbullet\
Final execution command:
\autorsevar{FINAL_EXECUTION_COMMAND}\par

\noindent\textbullet\
Agent summary: \autorsevar{AGENT_SUMMARY}\par

\autorserecordrole{Rating Scale}

\noindent\textbullet\
\textbf{0: Routine.}
No effective modeling changes, or only minor
hyperparameter adjustments.\par

\noindent\textbullet\
\textbf{1: Combinational.}
Reconfiguration of existing baseline components
without new mechanisms.\par

\noindent\textbullet\
\textbf{2: Exploratory.}
Addition of standard components, such as optimizers,
schedules, regularizers, features, or ensembles.\par

\noindent\textbullet\
\textbf{3: Hybrid.}
Non-trivial recombination of baseline components
with new mechanisms; simple module concatenation
is insufficient.\par

\noindent\textbullet\
\textbf{4: Transformational.}
Substantial changes to representations, learning
objectives, modalities, or task formulation;
scaling or ensembling alone is insufficient.\par

\autorserecordrole{Evidence Requirements}

Inspect the code diff together with the final command
to identify the active prediction pipeline.
Ignore unexecuted edits, abandoned alternatives,
and claims unsupported by code.
Do not infer novelty from performance gains or the
number of changed lines.
Assess departure from this baseline, rather than
absolute originality in the literature.
For scores 3 or 4, cite specific changed files,
functions, or diff excerpts and explain the mechanism
supporting the rating.
Flag ambiguous cases for review; leave the rating
unresolved if the active changes cannot be established.
Test scores must not be supplied or used.\par

\autorserecordrole{Required Output}

Return one integer score from 0 to 4, or unresolved
when evidence is insufficient; supporting code references;
a brief rationale; and a review flag.
Produce one rating per request.
Repeated ratings and aggregation are handled separately.\par

\end{tcolorbox}

\begin{table*}[!t]
\centering

\caption{Distribution of improvement strategies across agents.
\textnormal{Percentages are calculated over categorized
modification actions pooled across tasks and trials within
each model--harness combination.
$N$ denotes the number of categorized actions.
Bold indicates the most frequent category in each row.
Percentages may not sum to 100\% due to rounding.}}
\label{tab:strategy_distribution}

\footnotesize
\setlength{\tabcolsep}{3.5pt}
\renewcommand{\arraystretch}{1.05}

\resizebox{\textwidth}{!}{%
\begin{tabular}{@{}llrcccccc@{}}
\toprule

\multirow{2}{*}{Harness}
& \multirow{2}{*}{Model}
& \multirow{2}{*}{$N$}
& \multicolumn{6}{c}{Modification type (\%)} \\

\cmidrule(l){4-9}

& & &
Training & Model & Model & Feature & Parameter & Objective \\
& & & Strategy & Ensemble & Architecture & Modeling & Tuning & Function \\

\midrule

Claude Code & DeepSeek V4 Flash
& 396
& 8.6 & \textbf{61.9} & 4.0 & 4.3 & 18.7 & 2.5 \\

Claude Code & Claude Opus 4.8
& 405
& 3.0 & \textbf{83.2} & 1.5 & 1.7 & 9.9 & 0.7 \\

\addlinespace[2pt]

Codex & DeepSeek V4 Flash
& 552
& 6.5 & \textbf{55.6} & 3.6 & 5.8 & 26.6 & 1.8 \\

Codex & GPT-5.5
& 809
& 6.4 & 34.0 & 1.7 & 0.9 & \textbf{56.9} & 0.1 \\

\addlinespace[2pt]

Terminus-2 & DeepSeek V4 Flash
& 1,347
& 9.0 & \textbf{65.6} & 3.0 & 4.8 & 16.9 & 0.7 \\

Terminus-2 & GPT-5.5
& 617
& 1.8 & \textbf{65.3} & 2.3 & 7.3 & 23.0 & 0.3 \\

Terminus-2 & Claude Opus 4.8
& 299
& 8.0 & \textbf{53.2} & 9.0 & 4.0 & 22.7 & 3.0 \\

\midrule

\multicolumn{2}{l}{Overall}
& 4,425
& 6.6 & \textbf{59.0} & 3.1 & 4.2 & 26.2 & 1.0 \\

\bottomrule
\end{tabular}%
}

\end{table*}

\section{Task Descriptions}
\label{app:task_descriptions}

\subsection{Dataset Details}
Across the six benchmark tasks, the underlying evaluation data spans diverse real-world domains and formats, ranging from Amazon Beauty (22,363 users, 12,101 items with chronological sequence splits) and Kindle (100-candidate search-recommendation splits) to large-scale click logs including KuaiSAR v2 (1.46M test samples), Criteo-5M (4M/500K/500K train/val/test splits), and MovieLens Latest (over 2M total interaction samples).

\subsection{Baseline Reproduction and Correctness Verification}
\label{app:reproduction_verification}

To ensure that the initial baseline strategy $\pi_0$ across all six tasks provides a reliable, correct, and sound foundation for autonomous agent exploration, we carefully implemented and validated each codebase according to its original literature, official codebases, or trusted open-source community repositories.

\subsubsection{Strict Reproduction of Officially Released Baselines}
For tasks with official code releases, specifically \textbf{WuKong} and \textbf{HSTU}, we strictly reproduced the baselines following the exact architectural specifications, loss formulations, and hyperparameter configurations reported in their original publications. Under standard evaluation protocols on public datasets (MovieLens Latest and Amazon Beauty), our reproduced baselines achieved metric performances that strictly match or closely align with the official published results.

\subsubsection{Community-Guided Reproduction and Dataset Migration}
For tasks lacking official code releases but possessing community support (\textbf{TIGER}, \textbf{OneTrans}, and \textbf{RankMixer}), we constructed executable baselines by referencing reputable, high-quality open-source projects on GitHub (e.g., TIGER at \url{https://github.com/phonism/genrec}, OneTrans at \url{https://github.com/WestbrookLong/OneTrans_Pytorch}, and RankMixer at \url{https://github.com/samxiexs/RankMixer}). We rigorously migrated these implementations to standardized, accessible academic datasets (e.g., KuaiSAR v2 and Criteo-5M) while preserving their core algorithmic mechanics, feature interaction modules, and optimization frameworks.

\subsubsection{Best-Effort Independent Reproduction for Unreleased Baselines}
\textbf{IntSR} presented a unique challenge as neither official source code nor community implementations were publicly available. We performed a best-effort independent reproduction strictly based on the theoretical architecture, joint search-and-recommendation objectives, and scoring formulations described in the original paper. The implementation was built upon the UniSAR pipeline and adapted to the Amazon Kindle benchmark.

\subsubsection{Correctness and Evaluation Protocol Verification}
For all migrated and independently reproduced baselines, we thoroughly verified correctness by confirming stable optimization convergence, valid metric trajectories, and non-trivial competitive performance over standard reference models. Furthermore, all evaluation scripts (e.g., NDCG@10, AUC, and LogLoss) are implemented deterministically to guarantee fair and unbiased metric tracking across all agent iterations.

\section{Additional Experimental Results and Details}
\label{app:additional_results}

\subsection{Overall Distribution of Agent's Change}

Table~\ref{tab:strategy_distribution} provides the
detailed distribution of categorized modification
actions for each model--harness combination.
Actions are pooled across tasks and trials,
so the counts describe recorded research activity
rather than the number of runs.

\subsection{Segment Split Details}
To reconstruct structured action records from raw agent trajectories, we reuse existing action split points and reconnect relevant code modifications, execution outputs, and metric candidates based on these anchors. Specifically, the process consists of three parts: first, we locate the anchor of each action according to its tool call position in the trajectory; second, we associate any code edits or file modifications that occurred between the end of the previous action and the start of the current tool call as modification evidence for the current action; and finally, we collect the return results within the tool call and its subsequent interaction window (including direct outputs, background tasks, and terminal responses), extracting unvalidated metric candidates from them to fully restore the context of each trial.

\subsection{Complete results for all six tasks}

\begin{figure}[H]
    \centering
    \includegraphics[width=\columnwidth]{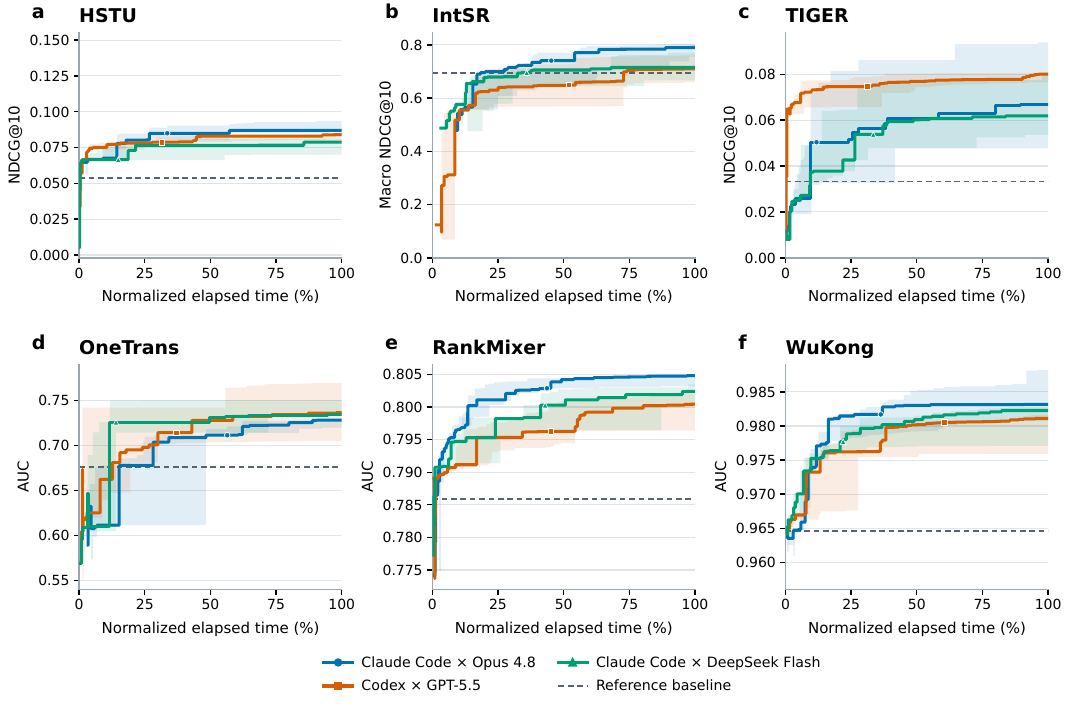}
    \caption{Optimization trajectories across the six recommendation tasks.}
    \label{fig:appendix_trajectories}
\end{figure}

\subsection{Case Studies}
\label{app:guidance_cases}

We trace supplied guidance to concrete implementations and
validation feedback.

\paragraph{HSTU: content information.}
The agent incorporates the provided 768-dimensional item
content embeddings through per-dimension standardization
and a bias-free linear projection. The projected features
are added to the learned item-ID embeddings in a table
shared by sequence input lookup and full-catalog scoring.
Content-fused models are retained in the final ensemble.

Validation NDCG@10 is 0.078877 with content fusion and
0.068080 without it, although the runs use early-stopping
patience values of 20 and 25, respectively.
A separate comparison adds a residual feed-forward stage
to the content-fused model, changing NDCG@10 from 0.080149
to 0.078877. The paired-user bootstrap 95\% confidence
interval for the difference is
$[-0.002898,\,0.000220]$.
The additional architectural component therefore shows
no clear incremental benefit in this configuration,
illustrating why structural complexity alone cannot
establish the usefulness of a research direction.

\paragraph{IntSR: training contrast.}
Training initially uses four negatives per example,
whereas evaluation uses 99.
The agent expands training candidates with positive items
from other examples in the minibatch, retains cross-entropy,
and masks target duplicates and observed-history negatives.

With minibatches of 128 and a target of 128 negatives,
the implementation fills 124 additional candidate slots
using in-batch items. Full minibatches therefore require
no additional popularity sampling, although that fallback
is implemented.
Models trained with this augmentation enter the final
ensemble. This documents an adaptation addressing the
training--evaluation contrast gap; its isolated contribution
to final performance is not measured.


\end{document}